\documentclass[notitlepage,letterpaper,showpacs,preprintnumbers,amsmath,nofootinbib,amssymb,superscriptaddress, hyperref,twocolumn]{revtex4-1}

\pdfoutput=1
\usepackage{amssymb,amsmath,latexsym,mathrsfs}
\usepackage[usenames,dvipsnames]{color}
\usepackage{graphicx} 
\usepackage{bm}
\usepackage{natbib}
\usepackage{todonotes}
\usepackage{color}
\usepackage[normalem]{ulem}
\usepackage[breaklinks,colorlinks,urlcolor=blue,citecolor=blue,linkcolor=blue]{hyperref}
\usepackage{epsfig,graphicx,verbatim,xspace,multirow,mathtools}
\usepackage{array}

\def\gsim{\mathrel{\raise.3ex\hbox{$>$\kern-.75em\lower1ex\hbox{$\sim$}}}}

\newcommand{\Tab}[1]{Table~\ref{#1}}

\begin{document}

\title{EDGES result versus CMB and low-redshift constraints on ionization histories}

\author{Samuel Witte}
\author{Pablo Villanueva-Domingo}
\author{Stefano Gariazzo}
\author{Olga Mena} 
\author{Sergio Palomares-Ruiz}
\affiliation{Instituto de F\'{\i}sica Corpuscular (IFIC), CSIC-Universitat de Val\`encia, \\
Apartado de Correos 22085,  E-46071, Spain}

\begin{abstract}
We examine the results from the Experiment to Detect the Global Epoch of Reionization Signature (EDGES), which has recently claimed the detection of a strong absorption in the 21~cm hyperfine transition line of neutral hydrogen, at redshifts demarcating the early stages of star formation. More concretely, we study the compatibility of the shape of the EDGES absorption profile, centered at a redshift of $z \sim 17.2$, with measurements of the reionization optical depth, the Gunn-Peterson optical depth, and Lyman-$\alpha$ emission from star-forming galaxies, for a variety of possible reionization models within the standard $\Lambda$CDM framework (that is, a Universe with a cosmological constant $\Lambda$ and cold dark matter CDM). When, conservatively, we only try to accommodate the location of the absorption dip, we identify a region in the parameter space of the astrophysical parameters that successfully explains all of the aforementioned observations. However, one of the most abnormal features of the EDGES measurement is the absorption amplitude, which is roughly a factor of two larger than the maximum allowed value in the $\Lambda$CDM framework. We point out that the simple considered astrophysical models that produce the largest absorption amplitudes are unable to explain the depth of the dip and of reproducing the observed shape of the absorption profile.
\end{abstract}
%%%%%%%%%%%%%%%%%%%%%%%%%%%%%%%%%%%%%%%%%%%%%%%%%%%%%
\maketitle
%%%%%%%%%%%%%%%%%%%%%%%%%%%%%%%%%%%%%%%%%%%%%%%%%%%%%

\section{Introduction}
The process known as \emph{reionization} identifies the cosmological epoch in which the first generation of galaxies appeared and began heating the surrounding medium. These galaxies emitted ultraviolet photons that ionized the neutral hydrogen, leading to the end of the so-called \emph{dark ages}. This process increased the number density of free electrons that could scatter off Cosmic Microwave Background (CMB) photons and hence, the reionization optical depth, $\tau$. The primary effect of an increase in the density of free electrons on the CMB temperature fluctuations is the suppression, by a factor $\exp(- 2 \tau)$, of the acoustic peaks at scales within the Hubble horizon at the reionization epoch. While this effect is highly degenerate with modifications of the amplitude of the primordial power spectrum, reionization processes also induce linear polarization on the CMB spectrum, leading to a “reionization  bump” at large scales. New results from the Planck collaboration in 2016 with an improved modeling and removal of unexplained systematics in the large angular polarization data~\cite{Aghanim:2016yuo} report a value of the optical depth $\tau$ smaller than in previous analyses~\cite{Ade:2015xua}. This new and refined Planck-CMB \texttt{SimLow} likelihood results in $\tau = 0.055 \pm 0.009$. Notice, however, that $\tau$ is an integrated quantity, and therefore provides a redshift-blind test of the reionization period. On the other hand, measurements of Lyman-$\alpha$ emission in star-forming galaxies and the Gunn-Peterson optical depth from bright quasars at low redshifts (indicating that reionization must have been completed by $z \sim 6$) can also constrain the reionization processes in the late Universe.

Experiments measuring the redshifted 21~cm line, arising from spin-flip transitions between the triplet and the ground singlet states in neutral hydrogen, offer unique probes to test both reionization and the dark ages of the Universe. The measured intensity of the 21~cm line depends on the ratio of the populations of the triplet and singlet states, which is expressed in terms of an effective excitation temperature, the so-called spin temperature $T_S$. The 21~cm signal can be measured either in emission or in absorption against the CMB, depending on whether the spin temperature is larger or smaller than that of the CMB. The differential brightness temperature, $\delta T_b$, in the Rayleigh-Jeans limit, is defined as
\begin{equation}
\delta T_b(\nu) = \frac{T_S - T_{\rm{CMB}}}{1 + z} (1 - e^{-\tau_{\nu_0}})~,
\label{eq:Tb}
\end{equation}
where $\tau_{\nu_0}$ is the optical depth of the intergalactic medium (IGM) for the 21~cm frequency $\nu_0 = 1420.4$~MHz. Given that the optical depth is small at all relevant redshifts, to first order in perturbation theory~\cite{Madau:1996cs, Furlanetto:2006jb, Pritchard:2011xb, Furlanetto:2015apc},
\begin{widetext}
\begin{equation}
\delta T_b(\nu) \simeq 27 \, x_\textrm{HI} \, (1 + \delta_b) \left( 1 - \frac{T_\textrm{CMB}}{T_S}\right) \left( \frac{1}{1+H^{-1} \partial v_r / \partial r} \right) \, \left( \frac{1+z}{10}\right)^{1/2} \left(\frac{0.15}{\Omega_m h^2} \right)^{1/2} \left( \frac{\Omega_b h^2}{0.023}\right)\,\textrm{mK} ~,
\label{eq:Tbdev}
\end{equation}
\end{widetext}
where $x_\textrm{HI}$ is the fraction of neutral hydrogen, $\delta_b$ is the baryon overdensity, $\Omega_b h^2$ and $\Omega_m h^2$ are the baryon and matter mass-energy densities, $H(z)$ is the Hubble function, and $\partial v_r / \partial r$ is the peculiar velocity gradient along the line of sight.

The Experiment to Detect the Global Epoch of Reionization Signature (EDGES)~\cite{Bowman:2012hf} has recently reported the measurement of an absorption profile centered at a frequency of $78 \pm 1$~MHz (i.e., at a redshift of $z \sim 17$) with an amplitude of $0.5^{+0.5}_{-0.2}$~K at $99\%$~CL~\cite{Bowman:2018yin}. This is about a factor of two larger than the maximum possible amplitude from predictions in standard $\Lambda$CDM scenarios. A number of studies have recently explored non-standard dark matter or dark energy physics to solve the issue~\cite{Munoz:2018pzp, McGaugh:2018ysb, Barkana:2018lgd, Barkana:2018qrx, Fraser:2018acy, Kang:2018qhi, Yang:2018gjd, Pospelov:2018kdh, Costa:2018aoy, Slatyer:2018aqg, Falkowski:2018qdj, Munoz:2018jwq} or to constrain these scenarios and other possible extensions of the standard cosmological picture~\cite{Fialkov:2018xre, Berlin:2018sjs, DAmico:2018sxd, Safarzadeh:2018hhg, Hill:2018lfx, Clark:2018ghm, Cheung:2018vww, Hektor:2018qqw,  Liu:2018uzy, Hirano:2018alc, Mitridate:2018iag, Mahdawi:2018euy}. The possibility of an enhanced radio background at early times has also been considered~\cite{Feng:2018rje, Ewall-Wice:2018bzf}. Here, we focus on the shape and redshift location of the absorption dip, studying its compatibility with CMB and low-redshift probes of the ionization fraction of the Universe within the standard $\Lambda$CDM paradigm.Indeed, we find a feasible region in the space of considered astrophysical parameters that provides a good fit to both the EDGES trough location and the different reionization observables. While none of the models explored here are capable of producing the absorption amplitude observed by EDGES, we also point out that combinations of astrophysical parameters leading to large absorption amplitudes (with maximal amplitudes $\delta T_b \gtrsim -280$~mK) cannot properly reproduce the shape of the absorption profile, where the shape of the profile is defined by the full-width at half maximum (FWHM) and the flatness in the proximity of maximum. Conversely, for models able to reproduce the shape and location of the absorption dip, the amplitude is significantly small ($\delta T_b \gtrsim -30$~mK). These results may point to an inconsistency in the observations when interpreted within the standard $\Lambda$CDM scenario, although this might not be necessarily the case and a rapid evolution of fundamental inputs of galaxy formation models at $z \gtrsim 10$ could also explain the EDGES results~\cite{Mirocha:2018cih}.

The structure of the paper is as follows. We present in Section~\ref{sec:reio} the methodology and the measurements considered in our statistical analyses. We discuss our results in Section~\ref{sec:results} and conclude in Section~\ref{sec:conclusions}.

\section{Methodology}
\label{sec:reio}

In this section we describe the simulation techniques followed to compute the averaged ionized fraction and the brightness temperature as a function of redshift. We also specify the reionization parameters we allow to vary and the cosmological/astrophysical observations used for our statistical analyses.

\subsection{Simulation and astrophysical parameters of the ionization history}

We make use of the publicly available code {\tt 21cmFast}~\cite{Mesinger:2010ne}, which, by means of a semi-analytic approach, generates simulations of the density, peculiar velocity, halo and ionization fields, as well as the brightness temperature.

The total ionized fraction $\bar{x}_i$ has two contributions. The major one is coming from the fully ionized HII\footnote{Note that HI and HII refer to neural and ionized hydrogen, respectively.} regions $Q_{\rm HII}$, while the sub-dominant one comes from the averaged ionized fraction of the neutral IGM. Including both contributions allows one to express the total $\bar{x}_i$ as~\cite{Mesinger:2012ys}
\begin{equation}
\label{eq:xetot}
 \bar{x}_i \simeq Q_{\rm HII}+(1-Q_{\rm HII}) \, x_e ~.
\end{equation}
The covering factor of the fully ionized regions is given by
\begin{equation}
Q_{\rm HII} = \frac{ \zeta_{\rm UV} \, f_{\rm coll}(>M_{\rm vir}^{\rm min})}{1-x_e} ~, 
\label{eq:HII} 
\end{equation} 
where $\zeta_{\rm UV}$ is the UV ionization efficiency, which is one of the astrophysical parameters varied in our analyses (see below), $f_{\rm coll}(>M_{\rm vir}^{\rm min})$ refers to the collapsed mass fraction into halos above a given threshold mass $M>M_{\rm vir}^{\rm min}$, and we use the default halo mass function in {\tt 21cmFast}~\cite{Sheth:1999mn, Sheth:1999su, Sheth:2001dp}. Once the function $\bar{x}_i $ is known, one can compute the reionization optical depth $\tau$:
\begin{equation}
\label{eq:tau} 
\tau=\sigma_T\int \bar{x}_i \, n_b \, dl ~,
\end{equation}
where $n_b$ is the baryon number density, $\sigma_T$ is the Thomson cross section and $dl$ is the proper distance along the line of sight. 

The astrophysical parameters governing the reionization and 21~cm signals that are allowed to vary in our analyses are: 
the UV ionization efficiency, $\zeta_{\rm UV}$;
the minimum virial temperature, $T_{\rm vir}^{\rm min}$;
the X-ray efficiency, $\zeta_{\rm X}$, which indicates the number of X-ray photons per solar mass in stars;
and, finally, the number of photons per stellar baryon between Lyman-$\alpha$ and the Lyman limit, $N_\alpha$.

\begin{figure*}[t]
	\includegraphics[width=\textwidth]{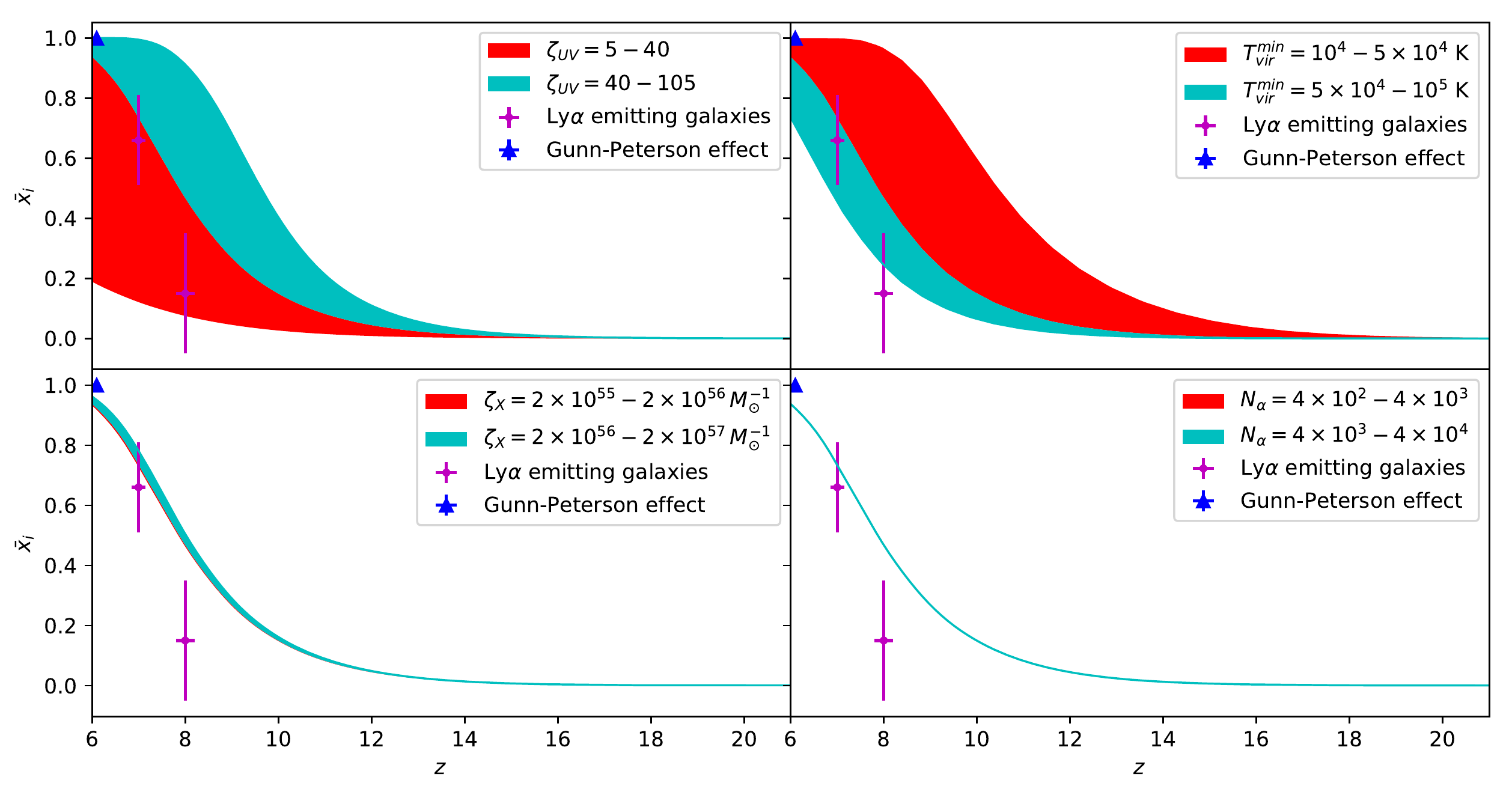}
	\caption{Evolution of the total ionized fraction of the neutral IGM, $\bar{x}_i$, as a function of redshift. The top/bottom left (right) plot depicts the results when $\zeta_{\rm UV}$/$\zeta_{\rm X}$ ($T_{\rm vir}^{\rm min}$/$N_\alpha$) is varied. When fixing their values, the astrophysical parameters are taken as $\zeta_{\rm UV} = 40$, $T_{\rm vir}^{\rm min} = 5 \times 10^4$~K, $\zeta_{\rm X} = 2\times 10^{56} \, M_\odot^{-1}$ and $N_\alpha = 4000$. The dots represent measurements of reionization observables~\cite{Fan:2005es, Schenker:2014tda, Bouwens:2015vha}.}
	\label{fig:xe_z}
\end{figure*}

The UV ionizing efficiency $\zeta_{\rm UV}$ is defined as the product of the fraction of baryons that form stars, the number of ionizing photons emitted per stellar baryon and the fraction of them that escape their host galaxy (see, e.g., Ref.~\cite{Mesinger:2012ys}), and it is assumed to be constant with redshift. Since it is closely related to the ionization fraction in the IGM (see Eq.~\eqref{eq:HII}), its value is crucial in the redshift dependence of $\bar{x}_i$. It is varied between $5$ and $105$~\cite{Greig:2015qca,Lopez-Honorez:2017csg}.
The minimum virial halo mass, which corresponds to the threshold mass for halos to host star-forming galaxies $M_{\rm vir}^{\rm min}$, can be related to $T_{\rm vir}^{\rm min}$ as~\cite{Barkana:2000fd}
\begin{equation}
\label{eq:mminT}
M_{\rm vir}^{\rm min} (z) \simeq 10^8 \left(\frac{T_{\rm vir}^{\rm min}}{2 \times 10^4 \, {\rm K}} \right)^{3/2} \left(\frac{1+z}{10}\right)^{-3/2} M_\odot ~.
\end{equation}
The minimum virial temperature is varied in the range $10^4$~K -- $10^5$~K (see, e.g., Refs~\cite{Evrard:1990fu, Blanchard:1992, Tegmark:1996yt, Haiman:1999mn, Ciardi:1999mx, Mesinger:2012ys, Greig:2015qca} and Refs.~\cite{Lopez-Honorez:2016sur, Villanueva-Domingo:2017lae, Escudero:2018thh}). A larger value of the minimum virial temperature implies a late star formation period, delaying the typical 21~cm signature as well as the overall reionization process.
%The UV ionization efficiency $\zeta_{\rm UV}$ is varied between $5$ and $105$, and since it is closely related to the ionization fraction in the IGM (see Eq.~\eqref{eq:HII}), its value is crucial in the redshift dependence of $\bar{x}_i$.
The X-ray heating efficiency, $\zeta_{\rm X}$, is varied from $2\times 10^{55} \, M_\odot^{-1}$ to $2\times 10^{57} \, M_\odot^{-1}$~\cite{Valdes:2012zv,Mesinger:2012ys,Christian:2013gma,Pacucci:2014wwa,Lopez-Honorez:2016sur}, corresponding to $N_{\rm X} \simeq 0.02$ and $N_{\rm X} \simeq 2$ X-ray photons per stellar baryon, respectively. This parameter affects both $\bar{x}_i$ and $\delta T_b$. Indeed, a large (small) value of $\zeta_{\rm X}$ implies an increase (decrease) in the gas kinetic temperature at higher redshifts, reducing (increasing) the amplitude of the dip in $\delta T_b$ and shifting its location to larger (lower) redshifts (see, e.g., Ref.~\cite{Mesinger:2012ys}). 
Finally, the parameter $N_\alpha$, whose default value in {\tt 21cmFast} is 4400 ionizing photons per stellar baryon in Pop-II stars, is varied within the range ($4 \times 10^2$, $4 \times 10^4$)~\cite{Barkana:2004vb}. We summarize the aforementioned parameters and their allowed ranges in \Tab{tab:params}.

\begin{table}
\setlength{\extrarowheight}{3pt}
\begin{tabular}{|c|c|}
\hline\hline
Parameter & Range  \\ \hline
$\zeta_{\rm UV}$ & $5 - 105$ \\ \hline
$\zeta_{\rm X}$ [$M_{\odot}^{-1}$]& $2\times10^{55}-2\times10^{57}$ \\ \hline
$T_{\rm vir}^{\rm min}$ [K] & $10^4 - 10^5$ \\ \hline
$N_\alpha$ & $4\times10^2-4\times10^3$ \\ \hline \hline
\end{tabular}
\caption{\label{tab:params} List of astrophysical parameters varied in this analysis and the range over which the parameters are restricted to vary. }
\end{table}

Figure~\ref{fig:xe_z} shows the evolution of the total ionized fraction of the neutral IGM, $\bar{x}_i$, as a function of redshift, together with the measurements used in this analysis (see Sec.~\ref{sec:data}). In each of the panels one of the four parameters above described is varied within the ranges quoted previously, while the other parameters are kept fixed to the following values:
$\zeta_{\rm UV} = 40$,
$T_{\rm vir}^{\rm min} = 5 \times 10^4$~K,
$\zeta_{\rm X} = 2\times 10^{56} \, M_\odot^{-1}$
and $N_\alpha = 4000$.
The top-left (right) plot depicts the values of $\bar{x}_i$ when $\zeta_{\rm UV}$ ($T_{\rm vir}^{\rm min}$) is varied. Notice that the impact of these two parameters is very large in the redshift evolution of $\bar{x}_i$: a lower value of $\zeta_{\rm UV}$, which delays the reionization process, can always be compensated by a lower value of $T_{\rm vir}^{\rm min}$, which would require a lower threshold mass for halos to host star-forming galaxies and therefore, would shift star formation processes towards earlier periods.
On the other hand, the bottom-left (right) plot depicts the range of ionization histories when $\zeta_{\rm X}$ ($N_\alpha$) is varied. Notice that the changes on $\bar{x}_i$ are much milder than those induced by the two previous parameters. Nevertheless, they play a crucial role in the redshift evolution of $\delta T_b$, as we shall see in what follows.

\begin{figure*}[t]
	\includegraphics[width=\textwidth]{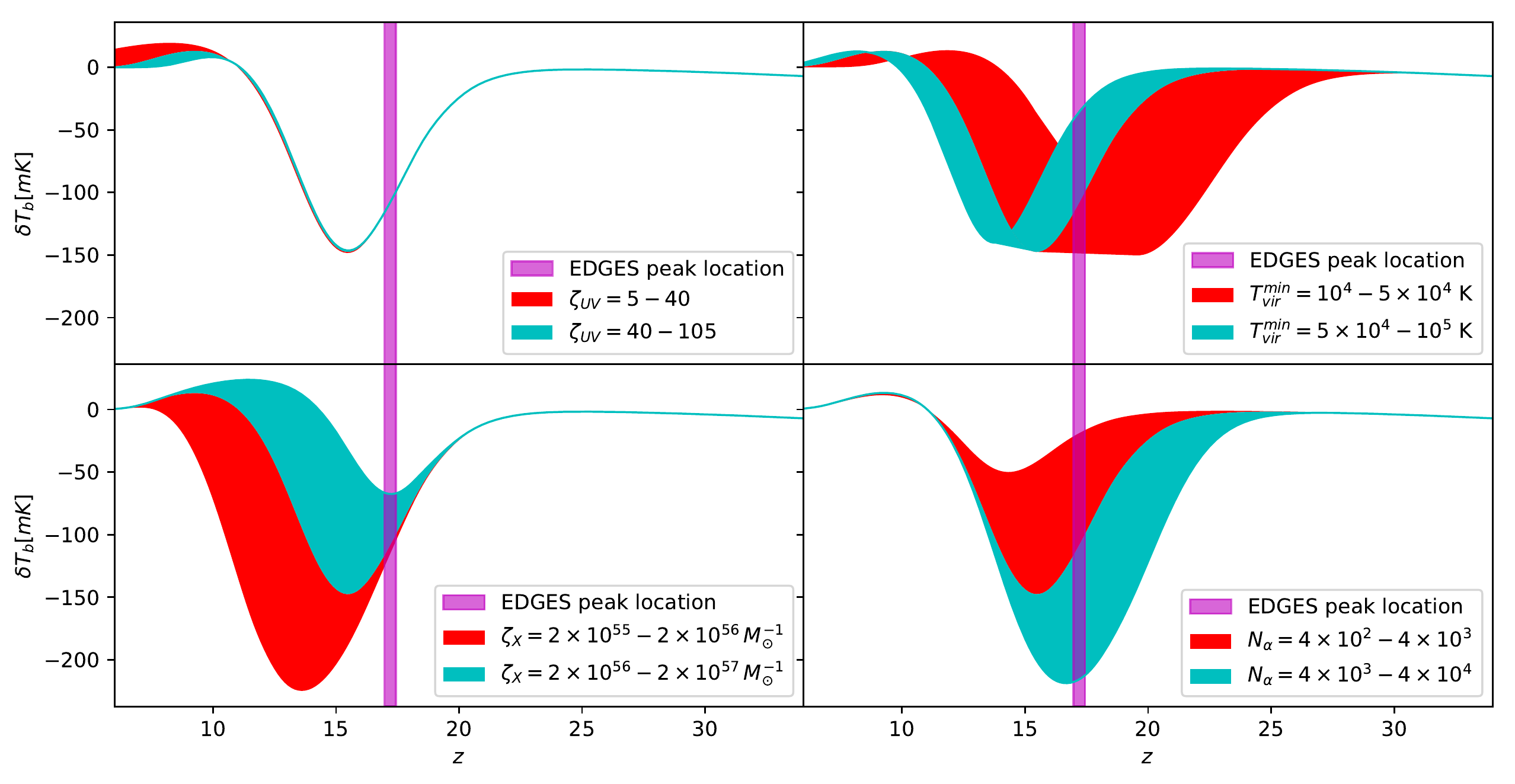}
	\caption{All-sky averaged $21$~cm differential brightness temperature as a function of redshift, varying one of the parameters considered in our analyses, and keeping fixed the remaining ones ($\zeta_{\rm UV} = 40$, $T_{\rm vir}^{\rm min} = 5 \times 10^4$~K, $\zeta_{\rm X} = 2 \times 10^{56} \, M_\odot^{-1}$ and $N_\alpha = 4000$). We indicate, with a purple vertical band, the position of the minimum of the absorption dip observed by EDGES, corresponding to $\nu = 78 \pm 1$~MHz.}
	\label{fig:Tb_z}
\end{figure*}

Before describing the effect of each of the four parameters on the evolution of the 21~cm sky-averaged signal, we briefly describe the typical behavior of the differential brightness temperature, $\delta T_b(z)$.
At redshifts $z \ge 300$, the spin temperature is coupled to the kinetic gas one ($T_K$) via collisions, and the latter itself is coupled to the CMB photon temperature via Compton scattering. As these three temperatures are almost identical, $\delta T_b \simeq 0$. Once CMB decouples, $T_S$ and $T_K$ are still coupled, but they evolve in redshift as $(1+z)^2$, while the CMB one evolves as $(1+z)$. Therefore, $\delta T_b <0$ and the 21~cm line is observed in absorption. As the IGM density decreases due to the Universe expansion, $T_S$ decouples from $T_K$, becoming closer to the CMB temperature, and therefore the differential brightness temperature vanishes at $z \simeq 30$, until redshifts at which the first sources become luminous. In this epoch, $T_S$ couples again to $T_K$ by means of resonant scattering of Lyman-$\alpha$ photons (this is the so-called \emph{Wouthuysen-Field effect}~\cite{Wouthuysen:1952,Field:1958,Hirata:2005mz}) and $\delta T_b$ is negative again, showing a characteristic absorption dip at $z \simeq 16-20$, presumably detected by EDGES. The typical amplitude of this dip is between $150$ and $200$~mK (see Fig.~\ref{fig:Tb_z}) and therefore, much shallower than the EDGES reported value. We shall comment on the implications of this absorption amplitude on our results later on. At redshifts $z \lesssim 16-20$, X-ray heating drives $\delta T_b$ towards less negative values so that it vanishes when the Universe is fully ionized, at $z \lesssim 10$.

In Fig.~\ref{fig:Tb_z}, we show the redshift evolution of the all-sky averaged 21~cm differential brightness temperature, varying one of the parameters considered in our analyses at a time, and keeping fixed the remaining ones to the same fiducial values used in Fig.~\ref{fig:xe_z}. The top-left panel of Fig.~\ref{fig:Tb_z} shows the $\delta T_b(z)$ function when the only varying parameter is the UV efficiency parameter. Notice that location and amplitude of the absorption dip and the overall shape of the 21~cm signal are barely affected, and only the reionization period, located at $z \simeq 6-10$, changes.
The top-right panel shows the changes in $\delta T_b(z)$ when the minimum virial temperature varies. In this case, while the absorption amplitude is unaffected, its location is significantly shifted: a low (high) value of $T_{\rm vir}^{\rm min}$ would imply an early (late) period of halos hosting star-forming galaxies. However, this shift could be compensated by a change in the X-ray efficiency or in $N_\alpha$. As shown in the bottom panels of Fig.~\ref{fig:Tb_z}, a lower value of either the X-ray heating efficiency or the $N_\alpha$ parameter could also shift the redshift absorption dip location and compensate the effect of a lower minimum virial temperature. These two parameters, $\zeta_X$ and $N_\alpha$, are also the ones that control the amplitude of the absorption signature in the 21~cm all-sky averaged brightness temperature. While a higher value of the X-ray efficiency would produce an earlier raise in the brightness temperature, diminishing the amplitude of the dip, the main effect of a larger value of $N_\alpha$ is the opposite (i.e., to produce a deeper trough in $\delta T_b(z)$), shifting its redshift location to earlier times. On the other hand, a lower value of $N_\alpha$ leads to a shallower absorption dip. Thus, a larger value of $N_\alpha$ could in principle be compensated by a higher X-ray efficiency. However, increasing both $\zeta_X$ and $N_\alpha$ shifts the redshift location of the absorption dip in the same direction (i.e., towards large redshifts) and a different value of $T_{\rm vir}^{\rm min}$ would also be required to not significantly change the shape of the all-sky 21~cm signal.\footnote{For a more complete discussion on the changes of the all-sky averaged 21~cm differential brightness temperature versus variations in the different astrophysical parameters, see Refs.~\cite{Mirocha:2015jra, Lopez-Honorez:2016sur}.} In principle, large values of $N_\alpha$ and small values of $\zeta_X$ would result in larger amplitudes. Nevertheless, the spin temperature cannot be lower than the gas temperature, and thus $\delta T_b$ can never be below $\sim - 280$~mK (see Eq.~\eqref{eq:Tbdev}).
Actually, it should be emphasized that the absorption saturates for the extreme values of $\zeta_X$ and $N_\alpha$ considered in this work, and none of the values considered produces a value within the range quoted by EDGES (i.e., $0.5^{+0.5}_{-0.2}$~K at $99\%$~CL~\cite{Bowman:2018yin}).

\subsection{Ionization history and differential brightness temperature measurements}
\label{sec:data}

The measurements we consider to constrain the ionization fraction of the Universe are:
\emph{(a)} the value of the reionization optical depth from the Planck-CMB \texttt{SimLow} likelihood results~\cite{Aghanim:2016yuo};
\emph{(b)} Gunn-Peterson optical depth at $z = 6.1$ from bright quasars~\cite{Fan:2005es}; and
\emph{(c)} Lyman-$\alpha$ emission in star-forming galaxies at $z \gtrsim 7$~\cite{Schenker:2014tda} (see also Ref.~\cite{Bouwens:2015vha}).

We compute $\chi^2$ functions for each of these data sets and compare the resulting constraints to those arising from the brightness temperature data. In order to evaluate the compatibility of the EDGES measurement with the numerical calculations of $\delta T_b(z)$ we adopt the following procedure:
for each choice of astrophysical parameters we extract 50 values of $\delta T_b(z)$ with the minimum between $\nu = 60$ and $99$ MHz (linearly spaced in frequency), subject each point to a random Gaussian fluctuation with a root mean square (rms) value of $0.087$~K (corresponding to the quoted rms errors determined without foreground modeling), and fit the resultant points using the flattened Gaussian absorption profile used by the EDGES collaboration:
\begin{equation}
\label{eq:fit_func}
\delta T_b(\nu) = - A \left(\frac{1 - e^{- \beta \, e^{B(\nu)}}}{1 - e^{- \beta}}\right) ~,
\end{equation}
with
\begin{equation}
B(\nu) \equiv \frac{4(\nu - \nu_0)^2}{\omega^2} \, \log\left[\frac{-1}{\beta}\log\left(\frac{1 + e^{- \beta}}{2}\right)\right] ~,
\end{equation}
where $A$ is the absorption amplitude, $\nu_0$ is the central frequency, $\omega$ is the FWHM and $\beta$ is the flattening factor. Repeating this procedure $\mathcal{O}(10^3)$ times for each choice of astrophysical parameters produces distributions for each of the best-fit parameters. For a parameter to be accepted as consistent with the EDGES measurement of the central frequency, we require that $\geq 10\%$ of the fits for a fixed choice of astrophysical parameters produce a central frequency within the quoted value of $78 \pm 1$~MHz. Note that our conclusions are relatively insensitive to this choice of $10\%$, and we have verified that taking instead thresholds of $5\%$ or $20\%$ lead to nearly identical results. Since the quoted measurements of EDGES are done \emph{a posteriori} using an arbitrary absorption profile, this procedure allows for a direct and meaningful comparison.

\section{Results}
\label{sec:results}

\begin{figure}[t]
	\includegraphics[width=0.49\textwidth]{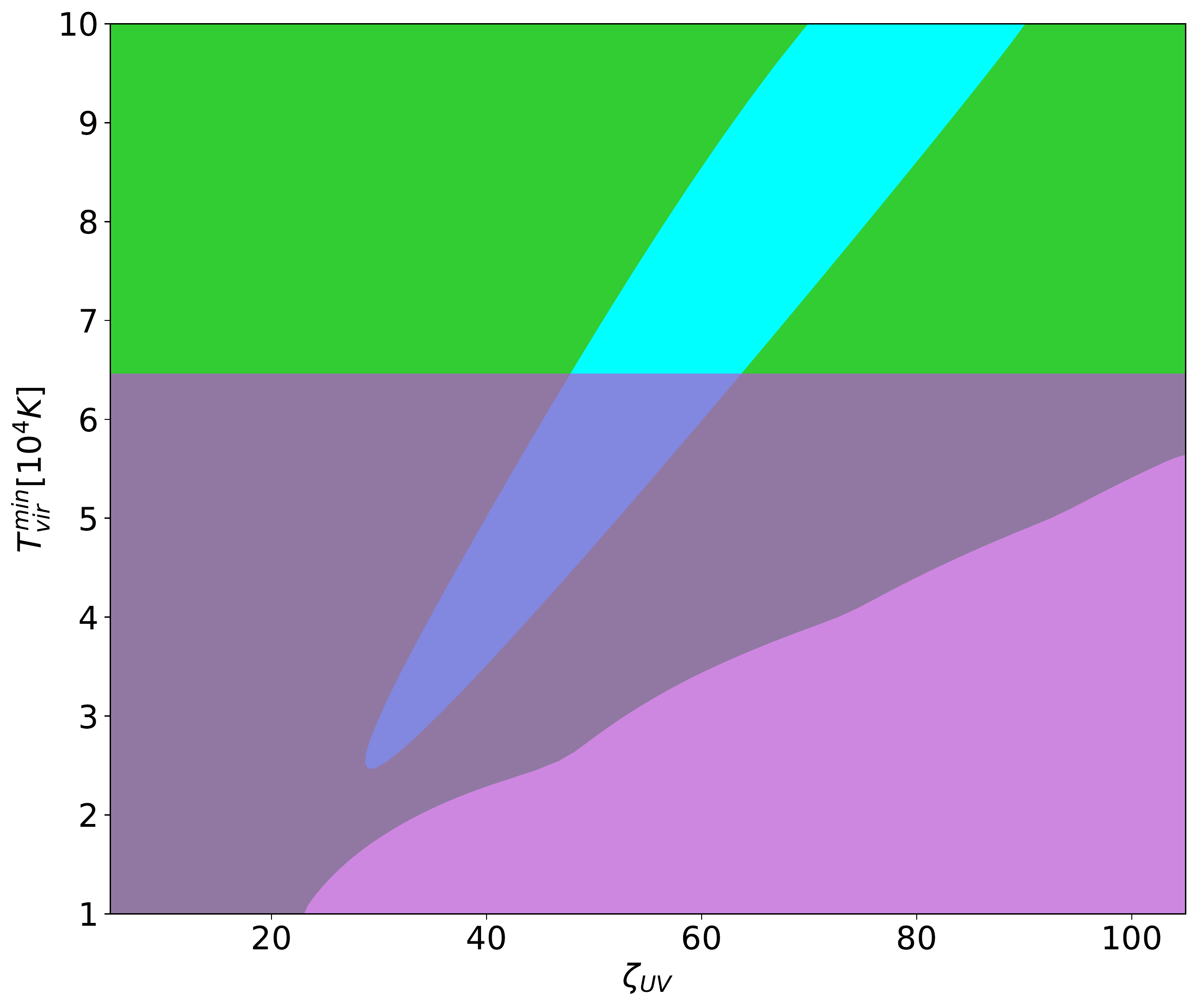}
	\caption{Contours allowed at $99\%$~CL in the ($\zeta_{\rm{UV}}, \, T_{\rm vir}^{\rm min}$) plane, based on CMB measurements of $\tau$ (green contours), $\bar{x}_i$ data (cyan contour), and the preferred area obtained by the position of the absorption dip in $\delta T_b$ extracted from EDGES (purple contour, that sets an upper bound on $T_{\rm vir}^{\rm min}$). See Sec.~\ref{sec:data} for details.}
	\label{fig:chi2}
\end{figure}

Figure~\ref{fig:chi2} shows the constraints on the ($\zeta_{\rm{UV}}, \, T_{\rm vir}^{\rm min}$) plane from the measurements described in the previous section, profiling over the other two parameters (i.e.,  for each point in $(\zeta_{\rm{UV}}, \, T_{\rm vir}^{\rm min}$)  we consider the maximum of the likelihood over all possible values of the other parameters) $\zeta_X$ and $N_\alpha$. 
Measurements of the reionization optical depth $\tau$ exclude (at $99\%$~CL) the region corresponding to large values of the UV ionization efficiency and low values of the minimum virial temperature, which would correspond to a very large value for $\tau$.
As expected, measurements of the ionization fraction, $\bar{x}_i$, result in a degenerate area in the ($\zeta_{\rm{UV}}, \, T_{\rm vir}^{\rm min}$) plane, since a lower value of $\zeta_{\rm{UV}}$ can always be compensated by a lower minimum virial temperature (see, e.g., Ref.~\cite{Lopez-Honorez:2017csg}). However, the observations related to the Gunn-Peterson effect exclude the region associated to very low values of the UV efficiency, as in this region reionization would likely be incomplete by $z \simeq 6$. The addition of the EDGES results, when only considering the frequency of the minimum of the absorption dip, provides a bound on the minimum virial temperature ($T_{\rm vir}^{\rm min} \lesssim 6.5 \times 10^4$~K), which is completely independent of $\zeta_{\rm{UV}}$, as expected (see the top left panel of Fig.~\ref{fig:Tb_z}). For larger values of $T_{\rm vir}^{\rm min}$, the onset of the absorption dip would occur at lower redshifts, in tension with the observed EDGES absorption window. Nevertheless, there exists a \emph{sweet spot}, in which the regions preferred by the ionization history of the Universe and the EDGES observations of the dip location overlap. Indeed, the fact that these measurements are mainly sensitive to different ranges of values of some of the astrophysical parameters (in particular $T_{\rm vir}^{\rm min}$) represents an interesting synergy which could help to disentangle reionization models. On the other hand, we caution that there are significant uncertainties in the description of the halo mass function and its redshift dependence and that all the results presented here are obtained assuming the Sheth and Tormen functional form~\cite{Sheth:1999mn, Sheth:1999su, Sheth:2001dp}, which is the default one used in {\tt 21cmFast}~\cite{Mesinger:2010ne}.

Thus far we have focused primarily on assessing the relative compatibility of the frequency of the absorption dip observed by EDGES with measurements from reionization. However, the absorption profile is also described in terms of its characteristic shape  (i.e., the width and flatness of the profile, characterized respectively by $\omega$ and $\beta$ in Eq.~\eqref{eq:fit_func}) and the depth of absorption (i.e., the amplitude $A$ in Eq.~\eqref{eq:fit_func}). The observed values at the $99\%$~CL of these parameters for the flattened Gaussian fit reported by EDGES correspond to: $\beta = 7^{+5}_{-3}$ MHz, ${\rm FWHM} = 19^{+4}_{-2}$ MHz, and an amplitude of $0.5^{+0.5}_{-0.2}$~K~\cite{Bowman:2018yin}. As previously mentioned, the absorption amplitude measured by EDGES is a factor of two or three larger than the maximum value we obtain in the considered models of reionization.

An additional question that remains is whether the properties of the absorption profile (i.e., the amplitude, shape, and location) are themselves self-consistent, and which features of the measurement can potentially yield compatibility with the reionization observables. In order to answer this question we use the generated fits for each set of astrophysical parameters described in Sec.~\ref{sec:data}. We begin by noting that the largest obtained absorption amplitude for the parameter space scanned here is approximately $\delta T_b\sim - 280$~mK, only slightly below the $99\%$~CL quoted by EDGES. As explained above, the astrophysical parameters giving rise to such amplitudes require small values of $\zeta_X$ and large values of $N_\alpha$. Nevertheless, the observed width of the profile, defined by the FWHM and $\beta$, can only be obtained within these reionization models if $\zeta_X$ is large and $N_\alpha$ small, illustrating a potential inconsistency of the observed absorption profile with a large absorption amplitude. It is also interesting to note that some of the reionization models providing consistency with the measured central frequency also provide consistency with the observed FWHM and $\beta$ (but not with the amplitude).
As expected, the extracted values of the FWHM and $\beta$ are insensitive to the value of $\zeta_{\rm{UV}}$.
As happens for the frequency of the absorption dip, smaller values of $T_{\rm vir}^{\rm min}$ are preferred to explain the FWHM and flatness of the observed profile, although large values of $T_{\rm vir}^{\rm min}$ could, in some cases, still be accommodated. Thus, there are sets of astrophysical parameters for which the shape and central frequency of the absorption profile maintain consistency with low redshift measurements of reionization, although the amplitude in those cases is much smaller than the one inferred from observations.

\begin{figure}[t]
	\includegraphics[width=0.5\textwidth]{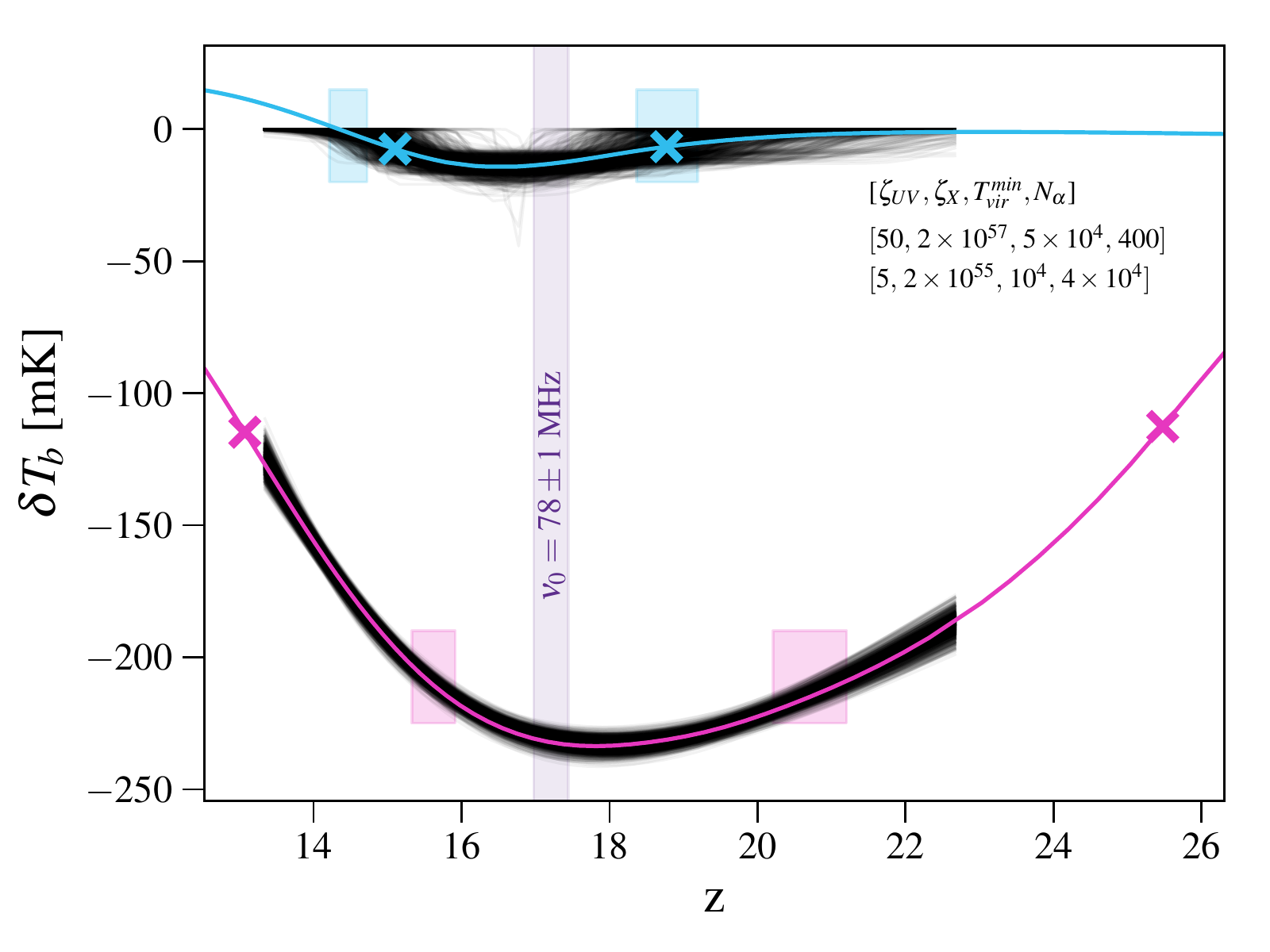}
	\caption{\label{fig:shapeanalysis}
		Numerical fits (black lines) of computed ionization histories (colored lines) using Eq.~\eqref{eq:fit_func} for one set of astrophysical parameters ($\zeta_{\rm{UV}} = 5$, $\zeta_X = 2 \times10^{55}$, $T_{\rm vir}^{\rm min} = 10^4$ and $N_\alpha = 4 \times 10^{4}$) producing large absorption amplitudes (magenta) and one set of astrophysical parameters ($\zeta_{\rm{UV}} = 50$, $\zeta_X = 2 \times 10^{57}, \, T_{\rm vir}^{\rm min} = 5 \times 10^4$ and $N_\alpha = 400$) that results in fits consistent with the observed FWHM and flattening parameter $\beta$ (blue). Vertical regions highlight preferred values of the observed central frequency (purple) and where the half maximum of each model should reside if the measured FWHM by EDGES is centered about the minimum $\delta T_b$ of each model (for comparison, we denote the true points of half maximum in each model with a cross, `x').
	}
\end{figure}

To further illustrate the potential inconsistency between the observed amplitude and the shape of the absorption profile, in Fig.~\ref{fig:shapeanalysis} we compare fits of the differential brightness temperature using Eq.~\eqref{eq:fit_func} for two sets of astrophysical parameters: one giving rise to a large amplitude ($\zeta_{\rm{UV}} = 5$, $\zeta_X = 2 \times 10^{55}$, $T_{\rm vir}^{\rm min} = 10^4$ and $N_\alpha = 4 \times 10^{4}$, in magenta); and the other one ($\zeta_{\rm{UV}} = 50$, $\zeta_X = 2 \times 10^{57}, \, T_{\rm vir}^{\rm min} = 5 \times 10^4$ and $N_\alpha = 400$, in blue) providing fits that are consistent with the observed shape of the absorption profile. Here, the thin black lines denote $\delta T_b(z)$ obtained using the flat Gaussian profiles fitted to the predicted values of $\delta T_b$ (shown with color lines), using a rms value of $0.087$ K. The purple vertical band in Fig.~\ref{fig:shapeanalysis} shows the preferred range of the central frequency of the absorption dip from EDGES. The blue (magenta) vertical bands indicate the upper and lower side of the interval defined by the EDGES FWHM, but centered on the minimum $\delta T_b$ from the blue (magenta) simulations. These redshifts must be compared with the true size of the FWHM interval from the same fits, whose extremes are denoted by crosses in the figure. It should be clear from Fig.~\ref{fig:shapeanalysis} that models with very low absorption amplitudes (and thus, badly inconsistent with the reported observations) can produce flatter profiles with FWHM and flatness that are more consistent with the fits obtained by EDGES than those models producing large amplitudes (yet, not large enough to be consistent with EDGES observations).

\section{Conclusions}
\label{sec:conclusions}

Recent observations by EDGES have provided the first measurement of the all-sky averaged differential brightness temperature $\delta T_b$ corresponding to the signal of the 21~cm hyperfine transition line of neutral hydrogen around the reionization epoch. Even if the amplitude of the measured absorption dip in the $\delta T_b(z)$ function lies below the maximum allowed value in standard $\Lambda$CDM cosmologies by about a factor of two, its location in redshift ($z \simeq 17.2$) does lie within the expected range.

In this work we have assessed the compatibility of the redshift behavior of the brightness temperature with measurements of the ionization history of the Universe within \emph{standard cosmological and astrophysical scenarios}, finding that there exists a region of parameter space in which both CMB reionization optical depth and low-redshift estimates of the ionization fraction of the Universe are perfectly compatible with a global 21~cm signature whose minimum is located at the values quoted by EDGES. However, we show that astrophysical models saturating the absorption and producing an amplitude maximally compatible with the observation of EDGES (albeit, still incompatible at the $99\%$ CL with the measured value of $0.5^{+0.5}_{-0.2}$~K) badly fail in producing the shape of the profile, pointing towards an additional discrepancy between the measurement and canonical astrophysical reionization scenarios interpreted within the context of the widely accepted $\Lambda$CDM model.

\section*{Acknowledgments}
SW, PVD and OM  are supported by the Spanish grants FPA2014-57816-P and FPA2017-85985-P of the MINECO and PROMETEO II/2014/050 of the Generalitat Valenciana.
SG is supported by the Spanish grants FPA2017-85216-P and PROMETEOII/2014/084 and from the European Union’s Horizon 2020 research and innovation programme under the Marie Sk{\l}odowska-Curie individual grant agreement No.\ 796941.
SPR is supported by a Ram\'on y Cajal contract, by the Spanish MINECO under grants FPA2017-84543-P and FPA2014-54459-P and by the Generalitat Valenciana under grant PROMETEOII/2014/049, and partially, by the Portuguese FCT through the CFTP-FCT Unit 777 (PEst-OE/FIS/UI0777/2013).
SW, PVD, OM and SPR also acknowledge support from the European Union's Horizon 2020 research and innovation program under the Marie Sk\l odowska-Curie grant agreements No.\ 690575 and 674896.
This work was also supported by the Spanish MINECO grant SEV-2014-0398.

%%%%%%%%%%%%%%%%%%%%%%%%%%%%%%%%%%%%%%%%%%%%%%%%%%%%%
\bibliography{biblio}
%%%%%%%%%%%%%%%%%%%%%%%%%%%%%%%%%%%%%%%%%%%%%%%%%%%%%

\end{document}